\documentclass[aps,prb,twocolumn,english,superscriptaddress,citeautoscript,preprintnumbers,amsmath,amssymb,floatfix,footinbib]{revtex4-2}
\usepackage{hyperref}
\hypersetup{colorlinks=true,linkcolor=red,citecolor=blue}
\usepackage{amsmath}
\usepackage{graphicx}
\usepackage{dcolumn}
\usepackage{float}
\usepackage{color}

\usepackage [english]{babel} 
\usepackage [autostyle, english = american]{csquotes} 
\MakeOuterQuote{"} 

\begin{document}


\title{Ferromagnetism and Metal-Insulator transition in F-doped LaMnO$_3$}		
\author{Ekta Yadav}
\thanks{E.Y and P.G contributed equally to this work.}
\affiliation{Department of Physics, Indian Institute of Technology, Kanpur 208016, India.}
\author{Pramod Ghising$^{\textcolor{red}{*}}$}
\altaffiliation[Present address: ]{Center for Integrated Nanostructure Physics (CINAP), Institute for Basic Science (IBS), Sungkyunkwan University, Suwon 16419, Republic of Korea.}
\affiliation{Department of Physics, Indian Institute of Technology, Kanpur 208016, India.}
\author{K. P. Rajeev}
\affiliation{Department of Physics, Indian Institute of Technology, Kanpur 208016, India.}
\author{Z. Hossain}
\email{zakir@iitk.ac.in}
\affiliation{Department of Physics, Indian Institute of Technology, Kanpur 208016, India.}
\affiliation{Institute of Low Temperature and Structure Research, Polish Academy of Sciences, Ok\'{o}lna 2, 50-422 Wroc\l{}aw, Poland.}
	
\begin{abstract}
We present our studies on polycrystalline samples of  fluorine doped LaMnO$_3$ (LaMnO$_{3-y}$F$_{y}$). LaMnO$_{2.5}$F$_{0.5}$ exhibits remarkable magnetic and electrical properties. It shows ferromagnetic and metallic behavior with a high Curie temperature of $\approx 239$ K and a high magnetoresistance of $-$64$\%$. This drastic change in magnetic properties in comparison to pure LaMnO$_3$ is ascribed to the presence of mixed-valence Mn ions driven by the F-doping at the O-sites, which enables double exchange (DE) in LMOF. Furthermore, the resistivity data exhibits two resistivity peaks at 239 K and 213 K, respectively. Our results point towards the possibility of  multiple double exchange hopping paths of two  distinct resistances existing simultaneously in the sample below 213 K.
	
\end{abstract}

\maketitle

\def\thefootnote{*}\footnotetext{These authors contributed equally to this work}\def\thefootnote{\arabic{footnote}}
\section{Introduction}
Perovskite manganites have been known to exhibit exotic phenomena such as colossal magnetoresistance (CMR), ferromagnetism, half-metallic behavior \cite{7, 9, 11} etc. Perovskites have the chemical formula ABO$_3$. In perovskite manganites the smaller B ions (transition metals) sit at the center of an oxygen octahedra BO$_6$ as shown in Fig. 1. The exotic properties of manganites are derived from the parent compound  LaMnO$_3$ (LMO), which is an insulating A-type antiferromagnet with a N\'{e}el temperature of 140 K \cite{1, 2, 3}. LMO being an antiferromagnetic insulator, is "interesting but useless", to quote Louis N\'{e}el from his Nobel prize speech. A common strategy to extract "useful" properties out of LMO is doping the A-sites with divalent atoms. A large number of these exciting phenomena results from doping the La-site of the parent compound LaMnO$_3$. Prominent examples include compounds such as La$_{1-x}$Sr$_x$MnO$_3$ (LSMO), La$_{1-x}$Pr$_{x}$Ca$_{0.375}$MnO$_3$ (LPCMO), La$_{1-x}$Ca$_{x}$MnO$_3$ (LCMO) etc., which exhibit rich phase diagrams at various dopant concentrations \cite{4, 5, 6, 2020-PRB-Ghising, 2019-PRB-Ghising}. It is observed that for a certain range of the value of the doping concentration $x$, we can induce ferromagnetism and metallic properties in LMO. In the case of LSMO, the range of $x$ is $0.2<x<0.5$ \cite{1994-JPSJ-Tokura, 1995-PRB-Urushibara}. 

Doping at the A-site or the  La-site with divalent ions results in a mixed valency of the Mn ions i.e. Mn$^{3+}$/Mn$^{4+}$ \cite{12, 13}. The Mn$^{3+}$/Mn$^{4+}$ ions occupy the B-site of the LMO lattice. Mixed valency of Mn is essential for the ferromagnetic and metallic behavior, which result from the double exchange (DE) hopping of electrons along the Mn$^{3+}-$O$-$Mn$^{4+}$  chains from one B-site to another B-site in the manganite lattice \cite{1951-PhysRev-Zener}. The DE hopping depends on the Mn$^{3+}-$O$-$Mn$^{4+}$ bond angle, $\theta$ \cite{1998-BMS-Bahadur}. Double exchange hopping is maximum for $\theta_\circ = 180^\circ$, and decreases as $\theta$ deviates from $\theta_\circ$. Consequently, ferromagnetic and metallic properties of manganites are very sensitive to distortions of the MnO$_6$ octahedra.

An alternate route to tune the properties of manganites is doping at the anion sites. Although La site doping of LMO has been studied extensively, there are very few reports on O-site doping of LMO. One possible way is to dope the O-site with fluorine \cite{25, 26}. Here, we present our studies on F doping of LMO sample. We observe, that on F doping, the LMO sample exhibits ferromagnetic and metallic behavior with high values of magnetoresistance (MR). The ferromagnetic properties are at par with La-site doped manganites. 
\begin{figure}[htp]
		\centering
		\includegraphics[width=0.9\linewidth]{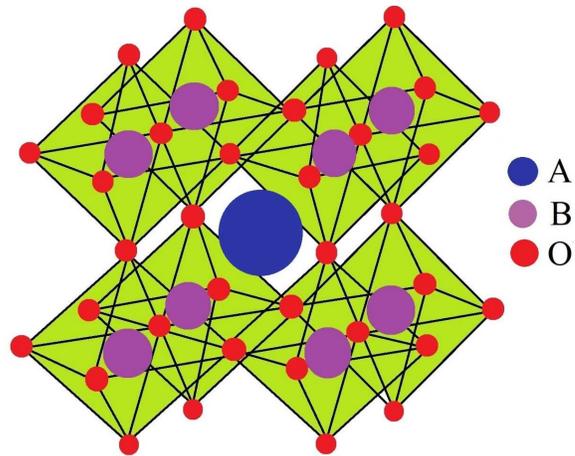}
		\caption{Structure of a perovskite oxide with the chemical formula ABO$_3$. The B-ion sits at the center of an oxygen octahedra BO$_6$.}	
\end{figure}
\section{Experimental Details}
\begin{figure}[htp]
		\centering
		\includegraphics[width=1.0\linewidth]{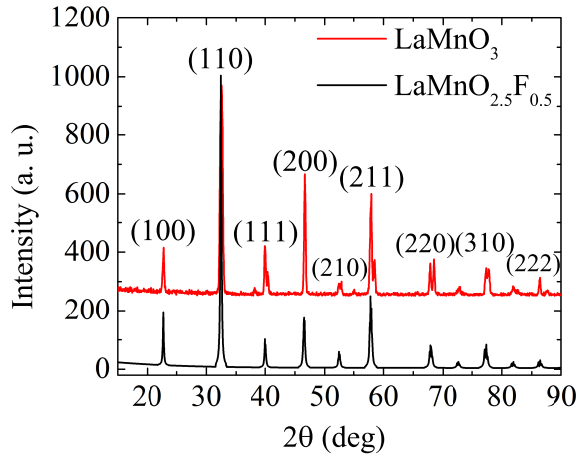}
		\caption{Powder X-ray diffraction result of LaMnO$_{3}$ and $y=$ 0.5 F-doped LaMnO$_3$ sample.}
\end{figure}
		
LaMnO$_{3-y}$F$_y$ (LMOF) samples (with $y=0.25, 0.50, 0.75$) were synthesized by the solid state reaction method. Powders of La$_2$O$_3$, Mn$_2$O$_3$ and LaF$_3$ were taken in stoichiometric ratios and were ground and mixed thoroughly using a mortar and pestle for 45 minutes. The ground powder was sintered in air for 24 hours at a fixed temperature of 600$^\circ$C, 700$^\circ$C, 800$^\circ$C, 900$^\circ$C, 1000$^\circ$C and 1100$^\circ$C. Before each sintering process, the powder was thoroughly ground in a mortar and pestle. Finally, the mixed powder was pressed into a pellet and annealed in air at 900$^\circ$C.
	
X-ray diffraction (XRD) and Energy-dispersive X-ray spectroscopy (EDX) was used for structural characterization and elemental analysis of the samples, respectively. XRD was performed using a Panalytical Expert Pro X-ray diffractometer with a Cu$-$K$_{\alpha1}$ radiation. JXA-8230 Electron Probe Microanalyzer (JEOL Ltd.) was used for determining the chemical composition. Iodometric titration was also used for determination of oxygen content. The electrical and magnetotransport measurements were carried out in a Quantum design physical property measurement system (PPMS). Standard four-probe technique was used for the electrical resistivity measurements. X-ray photoemission spectroscopy (XPS) measurement (using Al$-$K$_\alpha$ X-ray source, $h\nu=1486.6$ eV, PHI 5000 Versa Probe II, FEI Inc.) was carried out to determine the valence states of Mn and La in LMOF.
	
\begin{figure}[htp]
	\centering
	\includegraphics[width=1.0\linewidth]{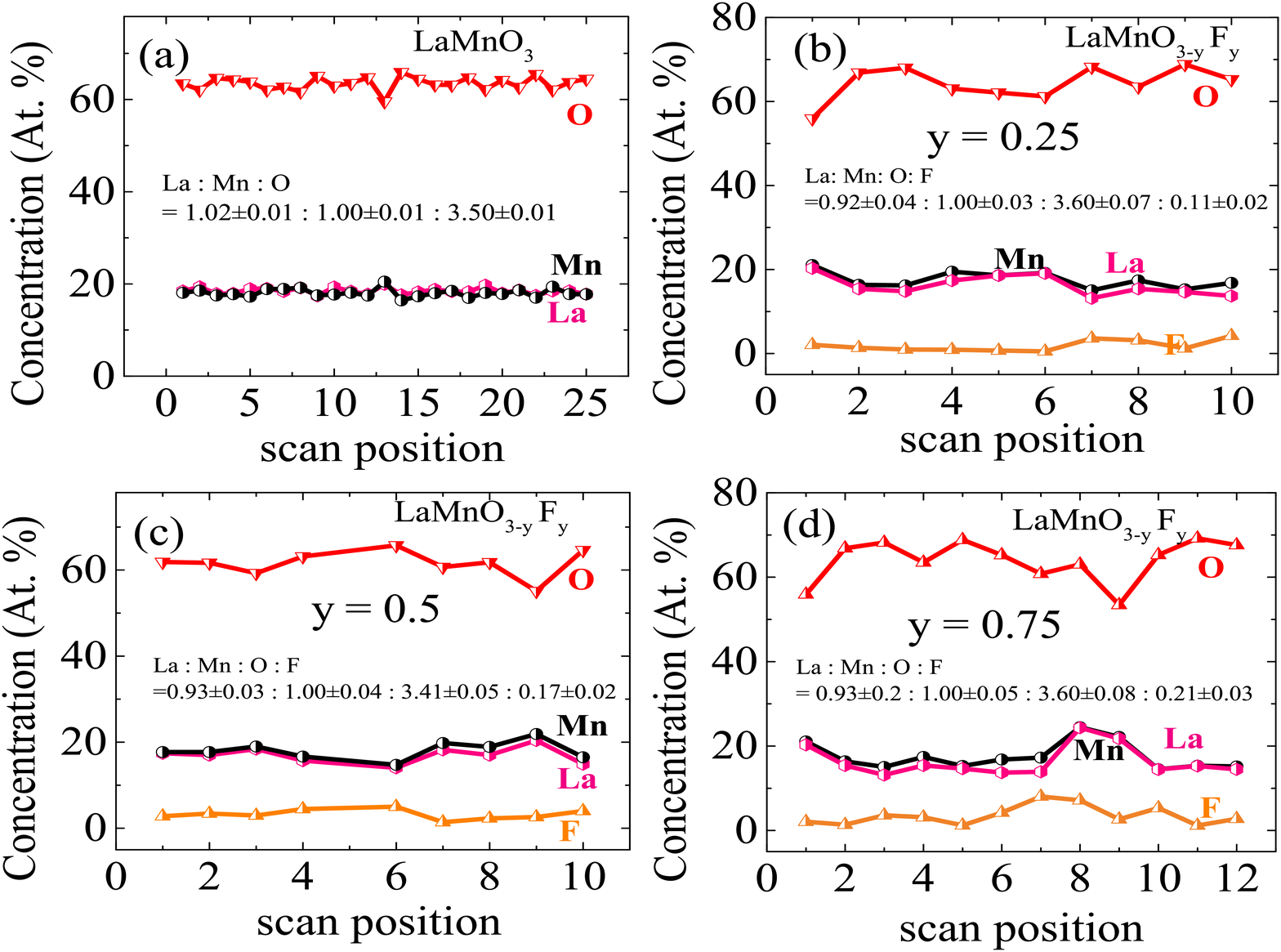}
	\caption{EDX average area points scan of various LMO samples: (a) undoped, (b) 0.25 F-doped, (c) 0.5 F-doped, (d) 0.75 F-doped.}	
\end{figure}

\section{RESULT AND DISCUSSION} 
The results of the XRD measurement carried out on the LMOF sample with $y=0.5$ is shown in Fig. 2(a) (unless stated explicitly, LMOF denotes F doping concentration of $y=0.5$ in the rest of the text). For comparison the XRD result of an LMO control sample is also shown in Fig. 2 (a) (red line). The XRD data reveals coinciding LMO and LMOF peaks indicating that LMOF also maintains the same perovskite structure as LMO. The absence of extra peaks in the XRD data confirms the absence of any secondary phases in the LMOF sample. The lattice parameters of LMO sample was determined to be $a = $ 0.5483 nm, $b = $ 0.7803 nm, $c = $ 0.5543 nm which are comparable to the values reported by others \cite{2017-PRB-Thygesen}. The lattice parameters of LMOF are $a = $ 0.5477 nm, $b = $ 0.7802 nm, $c = $ 0.5542 nm. The grain size of the different LMOF samples determined using Scherrer's formula are 29 nm, 39 nm and 32 nm for different F concentrations with $y=$ 0.25, $y=$ 0.50 and $y=$ 0.75, respectively. EDX measurements (at room temperature) on different regions of the LMOF sample show La deficiency. EDX data for undoped and F-doped LMO samples are shown in Fig. 3 along with atomic ratio of all elements. For undoped LMO, EDX [Fig. 3(a)] reveals La : Mn : O = 1.02 $\pm$ 0.01 : 1.00 $\pm$ 0.01 : 3.50 $\pm$ 0.01 with excess oxygen and with atomic concentration ($\pm$ standard error) La = 18.48 $\pm$ 0.46, Mn = 18.04 $\pm$ 0.78, O = 63.44 $\pm$ 1.99. The atomic ratio averaged over many point scans for different F-doped samples are depicted in the Fig. 3(b)-(d). For $y=$ 0.5 F-doped LMO it is La : Mn : O : F = 0.93 $\pm$ 0.03 : 1 $\pm$ 0.04 : 3.41 $\pm$ 0.05 : 0.17 $\pm$ 0.02 (with at. concentration La = 17.00 $\pm$ 0.60, Mn = 18.10 $\pm$ 0.65, F = 3.21 $\pm$ 0.35, O = 61.54 $\pm$ 0.97). For undoped LMO, at$\%$ of the cations are La=18.48 and Mn=18.04. Similarly, for F-doped LMO it is La=17.00 and Mn=18.10. It is clearly seen that the Mn content is more or less the same in both the doped and undoped LMO, whereas La content decreases in the F-doped sample. Thus, it is reasonable to normalize with respect to Mn rather than La. For all F-doped samples, fluorine concentration is seen to increase monotonically (0.11 $\pm$ 0.02, 0.17 $\pm$ 0.02, 0.21 $\pm$ 0.03) but is less than the intended value of \textit{y}=0.25, 0.5 and 0.75, respectively. EDX data shows considerably higher value of oxygen concentration for both undoped and F-doped LMO.

The low atomic number of oxygen makes it difficult for EDX measurements to make reliable estimate of its concentration in oxides. Therefore, we also performed iodometric titration\cite{New1,New2} to further determine the oxygen concentration in both doped and undoped LMO samples. Titration involves slowly adding a chemical substance to a reaction mixture till the chemical change is complete. For undoped LMO, iodometric titration yields a concentration of 3.21 with an excess oxygen of 0.21. The excess oxygen obtained with this method is much less than the EDX value of 0.5 for undoped LMO. Similarly, for $y=$ 0.5 F-doped LMO, it shows a concentration of 3.15, with excess oxygen of 0.15. This is again less than the excess oxygen value obtained from EDX for $y=$ 0.5 F-doped LMO (which is 0.41). Combining the EDX and iodometric titration we conclude that the undoped and doped samples have some excess oxygen and there is a systematic increase of F-content for different F-doped samples. The excess oxygen is due to sintering and annealing of the samples in air.

\begin{figure}[htp]
	\centering
	\includegraphics[width=1.0\linewidth]{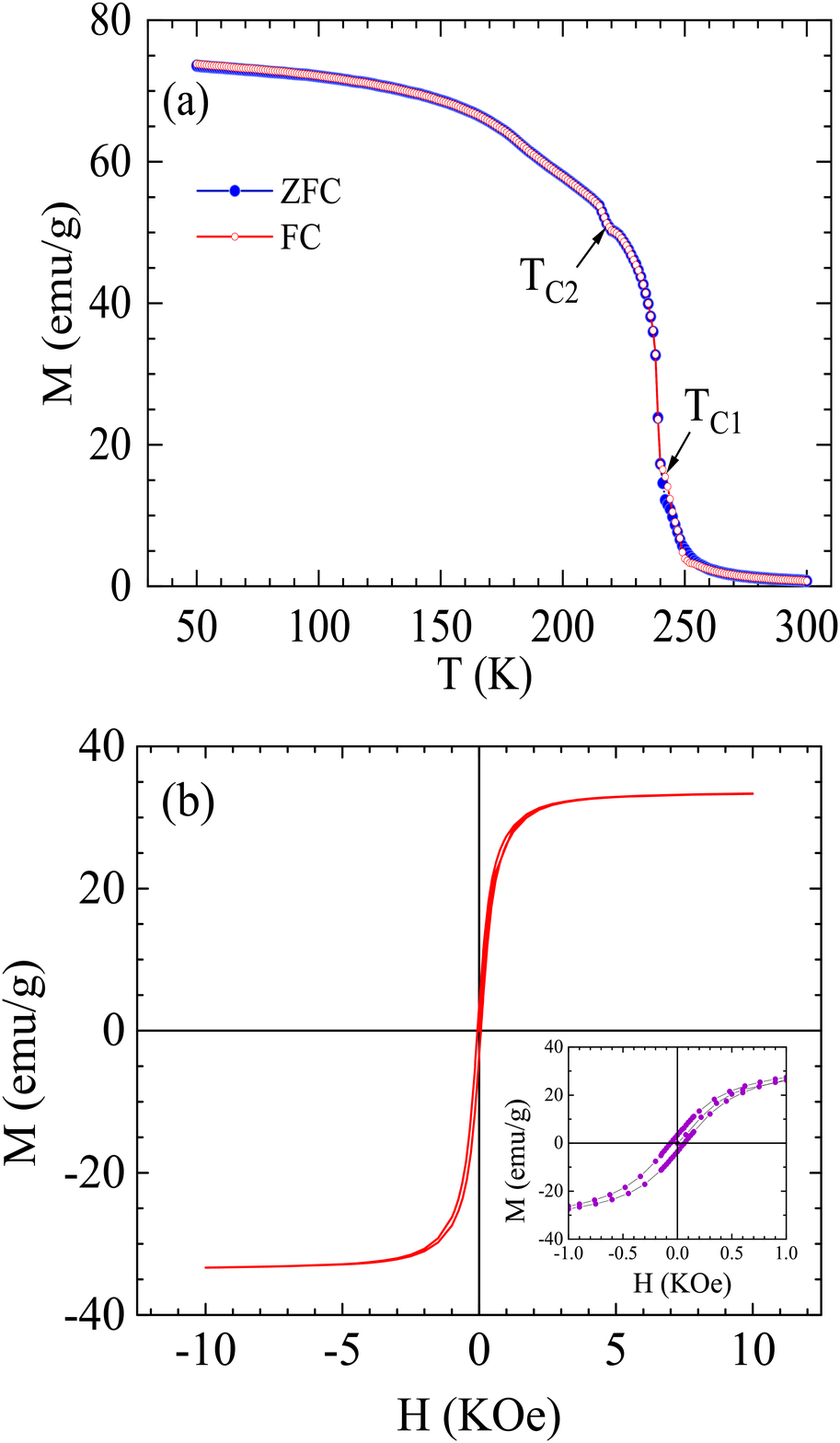}
	\caption{(a) Results of FC and ZFC-$M(T)$ measurements carried out on 0.5-F doped LMO. Two magnetic transitions in the $M(T)$ data are apparent at $T_{C1}=238$ K and $T_{C2}=217$ K. (b) $M(H)$ measurement of the 0.5-F doped LMO sample at 120 K ($< T_{C1}$, $T_{C2}$) exhibits a hysteresis loop, which confirms the ferromagnetic nature of 0.5-F doped LMO. The inset shows the blown up view of the central region of the hysteresis loop. The coercivity is 64 Oe.}	
\end{figure}

\begin{figure}[htp]
	\centering
	\includegraphics[width=1.0\linewidth]{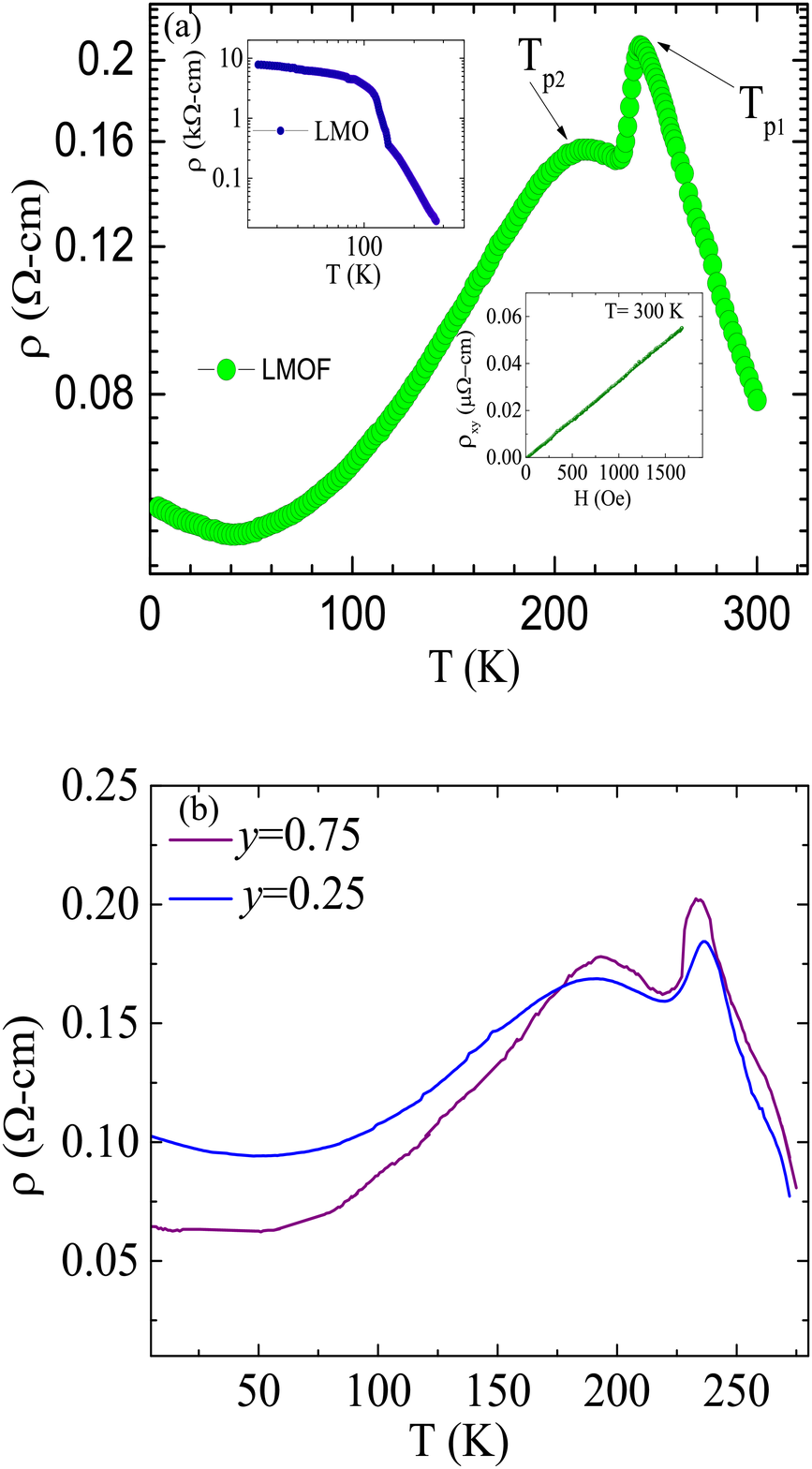}
	\caption{(a) Electrical transport measurement on the LMOF sample ($y=0.50$) exhibits metallic behavior with two distinct resistivity peaks at $T_{p1}=239$ K and $T_{p2}=213$ K. On the other hand, pure LMO sample exhibits insulating behavior as is observed from its resistivity data shown in the upper inset. The bottom inset shows the Hall measurement on the LMOF sample at 300 K. The positive slope in the Hall measurement indicates hole-type carriers in LMOF. (b) Resistivity measurement on samples with different F concentrations ($y=$0.25 and $y=$0.75) also exhibit metallic behavior with similar double peaks.}	
\end{figure}

\begin{figure}[htp]
	\centering
	\includegraphics[width=1.0\linewidth]{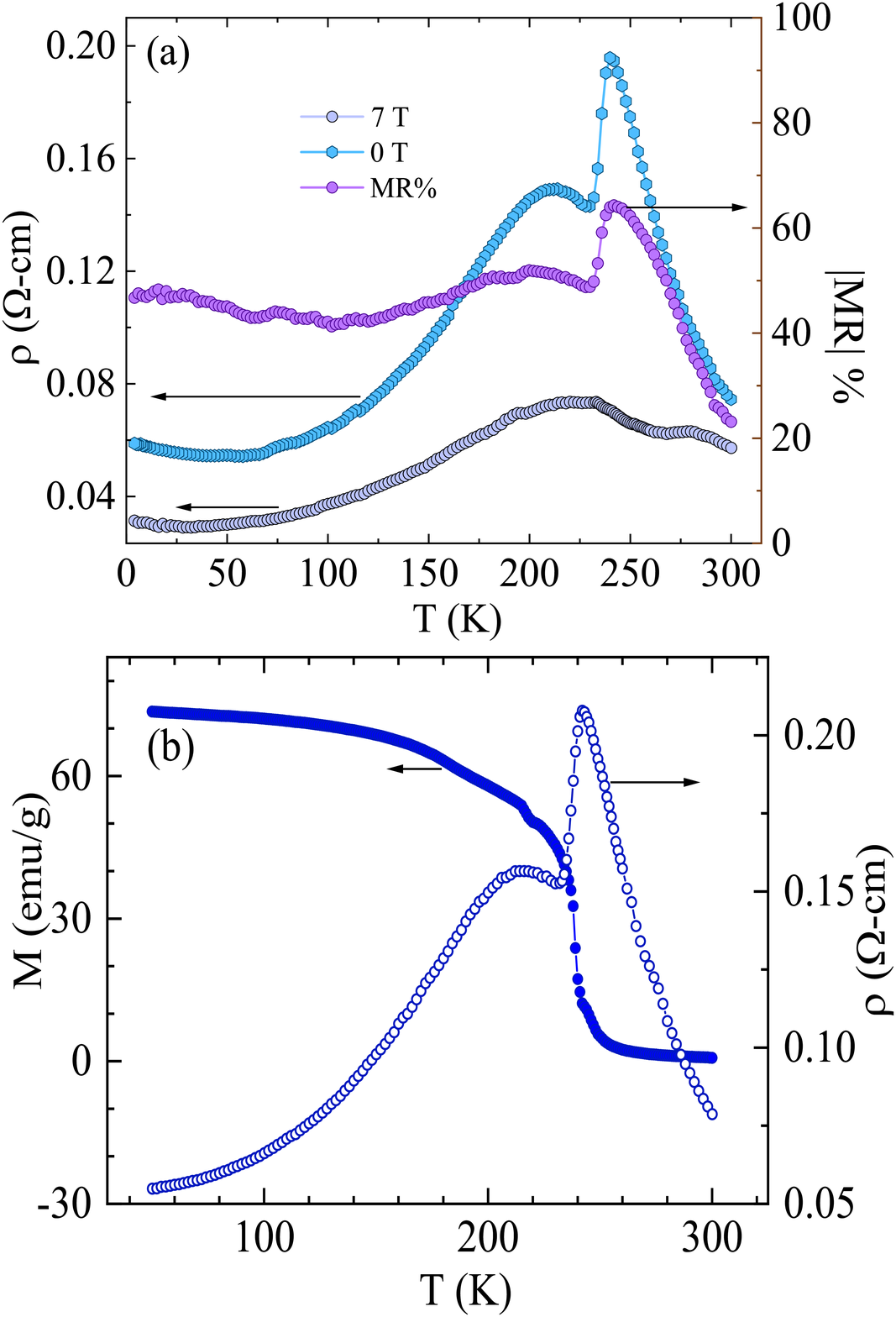}
	\caption{(a) Magnetic field dependent resistivity of the LMOF sample. At $T_{p1}$ and $T_{p2}$ a large magnetoresistance of $-$64$\%$ and $-$50$\%$ is observed. MR (shown on the right axis) is independent of temperature below  $T_{p2}$. (b) Comparison of the resistivity peaks with the magnetic transition temperature of LMOF. It is observed that the two magnetic transition temperatures in $M(T)$ correspond to the two resistivity peaks in $\rho(T)$. }	
\end{figure}

A very interesting feature of our LMOF sample is the observation of a ferromagnetic transition close to room temperature. Figure 4(a) displays the results of the field cooled (FC) and zero field cooled (ZFC) magnetization measurement as a function of temperature. For the FC measurement the sample was cooled in an applied field of 2000 Oe and the data acquisition was carried out during the heating cycle. The Curie temperature $T_{C1}$ of the ferromagnetic transition is 238 K. Another ferromagnetic transition is also observed in the $M(T)$ data at $T_{C2}=$ 217 K, indicating the presence of two distinct ferromagnetic exchanges in the LMOF sample. $M(H)$ measurement done at 120 K [shown in Fig. 4(b)] exhibits a hysteresis loop typical of ferromagnets, further confirming the ferromagnetic nature of LMOF below the Curie temperature.

To further probe the nature of ferromagnetism we carried out electrical transport measurements on the LMOF sample, the results of which are shown in Fig. 5(a). The sample exhibits a metallic behavior with a metal-insulator transition (MIT) at $T_{p1}=239$ K and another slightly broad resistivity peak at $T_{p2}=213$ K. It is to be noted that the undoped LMO control sample exhibits an insulating behavior without any MIT as can be seen in the upper inset of Fig. 5(a). Hall measurement carried out at 300 K on the LMOF sample reveals hole-type carriers, inferred from the positive slope of the $\rho_{xy}-H$ curve [shown in lower inset of Fig. 5(a)]. The carrier concentration was determined to be 0.18 $\times$ 10$^{22}$ cm$^{-3}$. For comparison, we also carried out resistivity measurements on two additional LaMnO$_{3-y}$F$_y$ samples with $y=$ 0.25 and $y=$ 0.75 . Their resistivity data also exhibit metallic behavior with similar double peaks as shown in Fig. 5(b). The field dependent resistivity data of LMOF is shown in Fig. 6(a). In the presence of a 7 T magnetic field both the resistivity peaks at $T_{p1}$ and $T_{p2}$ are suppressed and are also shifted to higher temperatures by an amount of 40 K and 22 K, respectively, suggestive of a similar magnetic origin of the two peaks. Moreover, a large magnetoresistance [MR$=\frac{{\rho (H) - \rho (0)}}{{\rho (0)}}$] of about $-$64$\%$ and $-$50$\%$ is also observed at $T_{p1}$ and $T_{p2}$, respectively. The MR below $T_{p2}$ show no significant change with temperature. Interestingly, the onset of the ferromagnetic transition $T_{C1}$ and the MIT temperature $T_{p1}$ coincide [shown in Fig. 6(b)]. This result strongly suggests that the ferromagnetic and the metallic properties of LMOF are related. One such example of concurrent ferromagnetic and metallic behavior in A-site doped LMO (such as LSMO, LCMO) results from double exchange mechanism \cite{1995-PRB-Urushibara, 1960-PhysRev-Gennes}, where the charge carriers are holes. Such DE mechanism in LMOF is only possible in the presence of mixed valence of Mn. Also, the resistivity peak $T_{p2}$ coincides with the ferromagnetic transition in the $M(T)$ data at $T_{C2}=217$ K, which further indicates that $T_{p2}$ is related to ferromagnetic properties of the sample.

The core level XPS spectra of the LMOF sample is shown in Fig. 7. Mn in LMOF exhibits a mixed valence state as is apparent from its XPS spectra displayed in Fig. 7(a). Analysis of the XPS spectra for Mn reveals a valence state of Mn$^{2+}$, Mn$^{3+}$ and Mn$^{4+}$. The Mn$^{2+}$, Mn$^{3+}$ and Mn$^{4+}$ peak positions occur at 641.0, 642.1, 643.6 and 653.0, 653.6, 654.6 for the 2$p_{3/2}$ and 2$p_{1/2}$ spin-orbit split peaks, respectively, which are similar to earlier reported values \cite{2015-SciRep-Huang, 1992-PSSa-Topfer}. The XPS spectra of La [shown in Fig. 7(b)] determines its valence state to be La$^{3+}$ \cite{2006-JAC-Mickevicius}. The deconvoluted La 3$d_{5/2}$ and 3$d_{3/2} $ show multiple peaks due to La binding with oxide and hydroxyl species and also final state effects. The peaks A, C, E and G correspond to La binding with oxides while the peaks B, D and F correspond to La binding with hydroxyl species \cite{2006-JAC-Mickevicius}. F 1$s$ and O 1$s$ XPS spectra are shown in Fig. 7(c) and (d), respectively. The additional peak at 531.5 eV in the O 1$s$ XPS spectra is ascribed to the formation of hydroxyl species \cite{2018-JPCM-Ghising}. 

Thus, F doping induces a mixed valence state of the Mn ions. Using resonant inelastic X-Ray Spectroscopy Orgiani \textit{et al.,} have established that Mn$^{2+}$ ions occupy the La site in LMO \cite{2010-PRB-Orgiani}, which probably explains the La deficiency in our LMOF sample. The origin of the Mn$^{2+}$ is due to the sample's tendency to maintain charge neutrality upon F doping. Mn$^{2+}$ ions occupying the La-sites is analogous to A-site doping of manganites by divalent atoms (such as in LSMO, LCMO etc.). It leads to the generation of Mn$^{4+}$ ions at the B (or the Mn) sites of LMOF. With the assumption that the La deficient sites (we deduce La deficiency of 0.07 per unit cell from EDX measurement) are occupied by Mn$^{2+}$, we obtain a doping level of 0.07 Mn$^{2+}$ ions per unit cell at the A-site. In analogy with A-site doped manganites (with divalent atoms such as Sr, Ca etc.), we obtain a carrier concentration of 0.07 holes per unit cell. Furthermore, Hall measurement at 300 K confirms hole-type carriers with a concentration of 0.18 $\times$ 10$^{22}$ cm$^{-3}$. Using the volume of the unit cell obtained from XRD data, we determine the carrier concentration to be 0.11 holes per unit cell, which is comparable with the EDX result. This result further supports Mn$^{2+}$ occupancy of the A-sites.

The presence of Mn$^{3+}$/Mn$^{4+}$ in LMOF does provide a fertile ground for the existence of double exchange, which would explain its metallic and ferromagnetic properties below $T_{C1}$. On the other hand, F$^-$ doping at the O$^{2-}$ sites will break the Mn$^{3+}-$O$-$Mn$^{4+}$ DE chain, which should degrade the metallic and ferromagnetic properties. The Mn ions occupying the A-sites have a considerably longer Mn$-$O bond lengths (2.7 \AA) as compared to when the Mn ions occupy the B-sites (1.9 \AA) \cite{2010-PRB-Orgiani}. Additionally, the occupation of the La sites by the smaller Mn$^{2+}$ ions would result in lattice distortions which would also lead to a decrease in DE hopping. Against this conflicting backdrop, the observed persisting ferromagnetism and metallic behavior of LMOF is very much intriguing.

\begin{figure}[htp]
	\centering
	\includegraphics[width=1.0\linewidth]{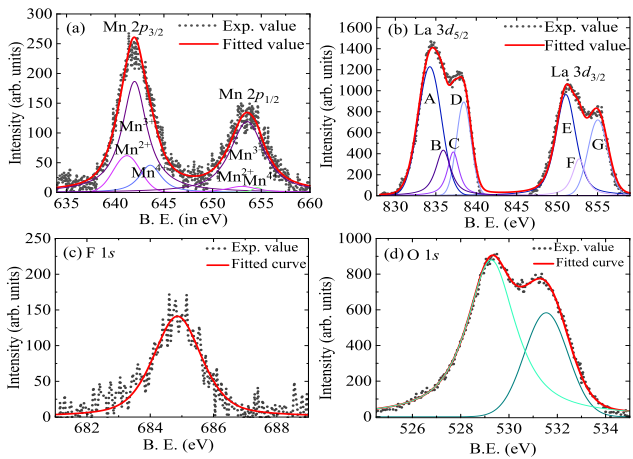}
	\caption{XPS spectrum of LMOF. The XPS data were fitted
		using a combination of Gaussian-Lorentzian peak shapes. (a) Mn 2$p$ spectrum with its two spin orbit doublet. The presence of Mn$^{2+}$, Mn$^{3+}$ and Mn$^{4+}$ peaks are observed in the XPS spectra. (b) La 3$d$ spectrum showing the two spin-orbit peaks, 3d$_{5/2}$ and 3d$_{3/2}$. Additional peaks due to La bonding with hydroxyl groups and final state effects are also observed. (c) F 1s spectrum (d)  O 1s XPS spectrum with an additional peak at 531.5 eV due to the formation of hydroxyl species.}
		 
\end{figure}

A fascinating feature of the LMOF sample is the presence of two resistivity peaks in $\rho(T)$ data at $T_{p1}$ and $T_{p2}$. Double peak in the resistivity of manganites have been reported earlier. Such double peaks were ascribed to the presence of excess oxygen in the manganites samples \cite{1997-PRB-Mandal}, grain size \cite{1999-JPCM-Zhang} and grain boundary effects \cite{1997-SSC-Ju, 2020-JaC-Verma}. The results of EDX and iodometric titration show oxygen excess in both undoped and F-doped LMO samples. However, double peaks in $\rho(T)$  is observed only in F-doped LMO. This confirms that the double peak feature originates exclusively from F-doping.
The peak at $T_{p2}$ that originate due to small polycrystal grain size exhibit low MR which is typically less than 20$\%$ as reported in literature \cite{1999-JPCM-Zhang}. However, the MR at $T_{p2}$ (at 7 T) is high in the LMOF sample, with a value of 50$\%$. The grain boundary effects manifest themselves as a gradual increase in MR as the temperature is lowered, with MR increasing to a value which is comparable to the MR value at $T_{p1}$ at the lowest temperatures \cite{2004-SSC-Yang}. No such MR behavior is observed in the LMOF sample [please see Fig. 6(a)].  MR is more or less independent of temperature below $T_{p2}$. The last two observations discount the possibility of grain size and grain boundary effects as the origin of double peak in the resistivity data. Another important feature to note is the presence of an additional magnetic transition in the $M(T)$ data at $T_{C2}$, which roughly corresponds to the second resistivity peak at $T_{p2}$ [see Fig. 6(b)]. Moreover, a large MR of $-$50$\%$ is observed at $T_{p2}$. These results indicate that the second resistivity peak is related to ferromagnetic properties of the sample.

\begin{figure} [htp]
	\centering
	\includegraphics[width=1.0\linewidth]{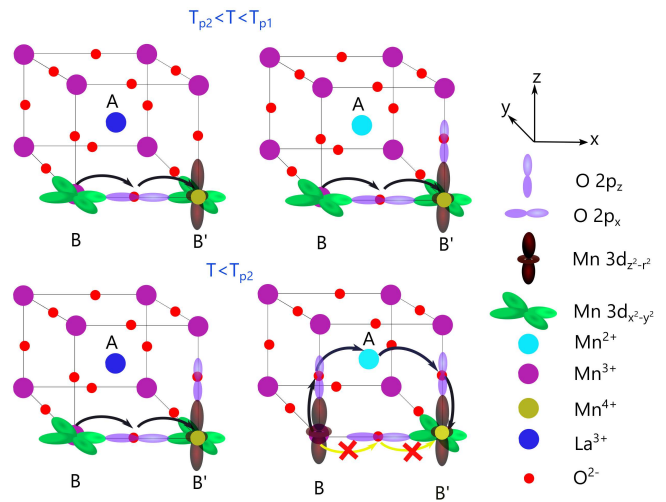}
	\caption{Schematic of the two DE hopping paths in LMOF. Top panel represents DE hopping in the temperature range $T_{p2}<T<T_{p1}$. Figure on the left shows a lattice with La$^{3+}$ at the A site, whereas on the right we have a lattice with Mn$^{2+}$ at the A-site. Hopping takes place by conventional DE along the low resistance path B$-$B$'$ for both the lattice. The bottom panel depicts the hopping mechanism in the temperature range $T<T_{p2}$. At $T_{p2}$, local orbital ordering of Mn$^{3+}$ $e_g$ (along z direction) takes place in the lattice with Mn$^{2+}$ at the A-site. As a result orbital overlap takes place along the $z$-direction. Consequently, DE hopping takes place along the high resistance path B$-$A$-$B$'$. In this lattice, conventional DE hopping along B$-$B$'$ is forbidden due to lack of orbital overlap along $x$-direction. On the contrary, for lattice with La$^{3+}$ at the A-site, there is no orbital ordering and hence conventional DE hopping along B$-$B$'$ survives.}
	
\end{figure}

The persistent ferromagnetism and metallic properties of LMOF may have its origin in the presence of Mn$^{2+}$ ions at the La site. Such occupancy of the A-site by divalent Mn$^{2+}$ would explain the hole-type carriers and the mixed-valent Mn$^{3+}$/Mn$^{4+}$ (in the LMOF sample) observed in the Hall and XPS measurements, respectively. 
The observed ferromagnetism may also originate from excess oxygen in LMOF. The excess oxygen can increase Mn$^{4+}$ concentration thereby facilitating double exchange. However, we do not observe ferromagnetism and metallic behavior in the undoped LMO sample with a similar oxygen excess. Thus, ferromagnetism in LMOF originates exclusively from F-doping.
In LMO containing Mn$^{2+}$, it has been shown that an additional double exchange hopping path via an intermediate Mn$^{2+}$ ion (Mn$^{3+}-$O$-$Mn$^{2+}-$O$-$Mn$^{4+}$) exists \cite{2010-PRB-Orgiani}, other than the conventional Mn$^{3+}$/Mn$^{4+}$ (Mn$^{3+}-$O$-$Mn$^{4+}$) hopping path. This DE hopping path involves electron hopping along B$-$A$-$B$'$ sites (B$'$ is a neighboring B site) in contrast to conventional DE, where the hopping sites are B$-$B$'$ in the manganite lattice (shown in Fig. 8). We suspect that the second peak in the resistivity data at $T_{p2}$ is on account of double exchange hopping along this additional path. The FM ordering at $T_{C2}$, further lends some credibility to our assumption. We could model the LMOF sample as consisting of some Mn-ions at B sites (Mn$^{3+}/$ Mn$^{4+}$) and some at A sites (Mn$^{2+}$), which exhibit ferromagnetic ordering at different temperatures $T_{C1}$ and $T_{C2}$, respectively. The ferromagnetic ordering of the A and B sites originate from DE hopping of electrons via the B$-$B$'$ and B$-$A$-$B$'$ hopping paths. Furthermore, the two DE paths exhibit different resistance. The higher resistance path include DE hopping via the Mn$^{2+}$ ions located at the A-sites. Due to the decreased angle ($<$180$^\circ$) between the Mn ions and the longer hopping path, it will have a higher resistance. Whereas the lower resistance path involves DE hopping via the conventional Mn$^{3+}-$O$-$Mn$^{4+}$ chains which occupy the B-sites with angle between the neighboring Mn ions close to 180$^\circ$. In the temperature range of $T_{p2}<T<T_{p1}$ the metallic behavior is due to DE hopping via the low resistance path. At $T_{p2}$ the Mn$^{2+}$ ions at the A-sites start to exhibit ferromagnetic order, which implies that a DE hopping via the high resistance path is also activated. One possible trigger for the activation of DE hopping along B$-$A$-$B$'$ is local orbital ordering of Mn$^{3+}$ $e_g$ orbitals. 
Local ordering of Mn$^{3+}$ $e_g$ orbitals have been observed in thin films of LSMO and LMO \cite{2011-PRB-Galdi, 2012-PRB-Aruta}. Here due to the presence of Mn$^{2+}$ at the A-sites, $e_g$ orbitals of Mn$^{3+}$ (at the B-sites) order along the $z$ direction (i.e. electrons occupy the $d_{z^2-r^2}$ orbital) \cite{2011-PRB-Galdi, 2012-PRB-Aruta}. Consequently, for those lattices where Mn$^{2+}$ ions occupy the A-site, hopping takes place via the $d_{z^2-r^2}$ orbital.
It is also interesting to note that this orbital ordering prevailed despite the presence of a tensile strain from the substrate on the LSMO film which would have otherwise preferred an in-plane (i.e. $d_{x^2-y^2}$) orbital ordering \cite{2012-PRB-Aruta}. Thus, such local orbital ordering of Mn$^{3+}$ is attributed to the presence of Mn$^{2+}$ at the A-sites. Such a local orbital ordering of Mn$^{3+}$ $e_g$ orbitals is the only possible scenario that favor hopping along B$-$A$-$B$'$ as shown in Fig. 8 (since DE hopping takes place only in the presence of orbital overlap between the participating Mn ions and the intermediate O ion). Mn$^{3+}$ being Jahn-Teller (JT) active, its $e_g$ orbitals ($d_{x^2-y^2}$ and $d_{z^2-r^2}$) are non degenerate, whereas Mn$^{2+}$ and Mn$^{4+}$ are JT inactive and hence their $e_g$ orbitals are degenerate. In the temperature range $T_{p2}<T<T_{p1}$, electrons occupy the lower energy $d_{x^2-y^2}$ in-plane orbital. As a result the orbital overlap is in the in-plane direction leading to conventional DE (as shown in top panel of Fig. 8). At $T_{p2}$ the Mn$^{3+}$ $e_g$ orbitals in the lattice containing Mn$^{2+}$ ions at the A-sites order locally along the out-of plane direction, i.e. the energy of the $d_{z^2-r^2}$ orbital is lowered. Consequently, orbital overlap with the O ions take place along the out-of plane direction as shown in the bottom panel of Fig. 8. This results in a DE hopping along the high resistance B$-$A$-$B$'$ path. Thus, below $T_{p2}$ the electron hopping is along the low resistance B$-$B$'$ and high resistance B$-$A$-$B$'$ paths in series. Such low resistance and high resistance paths in series leads to a re-emergent MIT in manganites \cite{2008-PRL-Ward}. The second resistivity upturn at $T_{p2}$ in the LMOF sample can be explained in terms of addition of the resistance of the two paths. Initially at $T_{p2}$, the resistivity of the sample increases due to the inclusion of the high resistance B$-$A$-$B$'$ path. Below $T_{p2}$, more number of hopping paths via Mn$^{2+}$ are activated. As a result, the resistance starts to decrease again after an initial increase at $T_{p2}$ as temperature is further lowered.

\section{Conclusion}

To conclude, we prepared flourine doped LaMnO$_{3}$ polycrystalline samples using solid state reaction method. The LMOF samples exhibit ferromagnetic and metallic properties which are similar to La-site doped manganites. The Curie temperature is 239 K and a high MR of $-$64$\%$ is observed. The ferromagnetic and metallic behavior is ascribed to DE mechanism. Furthermore, the resistivity data exhibits two resistivity peaks around the same temperature at which the $M(T)$ data exhibit two magnetic transitions. Our XPS studies show the presence of Mn$^{2+}$, Mn$^{3+}$ and Mn$^{4+}$ ions in the LMOF sample. The two peaks in the resistivity data are attributed to the presence of multiple DE hopping paths i.e. one along Mn$^{3+}-$O$-$Mn$^{4+}$ and another along Mn$^{3+}-$O$-$Mn$^{2+}-$O$-$Mn$^{4+}$. Our study shows that anion site doping of perovskite manganite can be a robust technique to explore new and exciting physics in perovskite manganites.

\section{Acknowledgement}
We gratefully acknowledge financial support from IIT Kanpur, India and Department of Science and Technology, India. ZH would like to acknowledge the Polish National Agency for Academic Exchange (NAWA) for Ulam Fellowship. We also thank Namrata Singh for helping us with iodometric titration.
\bibliography{Reference}
\end{document}